\documentstyle[11pt,graphics]{article}

\begin{document}

\begin{center}
{\large {\bf Semiclassical dressed states of two-level quantum systems
driven by nonresonant and/or strong laser fields\vspace*{0.6cm}\\}} A.
Santana, J. M. Gomez Llorente\footnote{%
e-mail: jmgomez@ull.es}, and V. Delgado\footnote{%
e-mail: vdelgado@ull.es} {\vspace*{.2cm}}\\{\it Departamento de F\'\i sica
Fundamental II,\\Universidad de La Laguna, 38205-La Laguna, Tenerife, Spain 
%\\e-mail: vdelgado@ull.es\\e-mail: jmgomez@ull.es
\vspace*{0.7cm}\\}
\end{center}

\begin{abstract}
Analytical expressions for the semiclassical dressed states and corresponding
quasienergies are obtained for a two-level quantum system driven by a
nonresonant and/or strong laser field in a coherent state. These expressions
are of first order in a proper perturbative expansion, and already contain
all the relevant physical information on the dynamical and spectroscopic
properties displayed by these systems under such particular conditions. The
influence of the laser field parameters on transition frequencies, selection
rules, and line intensities can be easily understood in terms of the
quasienergy-level diagram and the allowed transitions between the different
semiclassical dressed states.
\end{abstract}

\hspace{6. pt} PACS number(s): 42.50.-p, 42.50.Ct

\vspace{1.4 cm}

\begin{center}
{\large {\bf {I. INTRODUCTION\vspace{.1 cm}\\}}}
\end{center}

The driven two-level model has been extremely useful in the study of the
interaction of coherent light with atoms and molecules. Under resonant or
near resonant conditions ($\Delta _0/\omega _{{\rm {L}}}\simeq 1$; with $%
\Delta _0$ being the transition frequency and $\omega _{{\rm L}}$ the
driving field frequency) and not too strong laser fields ($\Omega _0/\omega
_{{\rm {L}}}\ll 1$; with $\Omega _0$ being the Rabi frequency) one can
invoke the rotating wave approximation and obtain an analytical description
which is valid in the lowest order of the coupling constant. This approach,
which is essentially the semiclassical Jaynes-Cummings model [\ref{Jaynes}],
corresponds to the regime most commonly encountered when considering atoms
in the presence of a laser field.

The two-level model has also been used to describe interaction of matter
with stronger or nonresonant laser fields. A first thorough study of the
relevant effects of a strong oscillating field on a two-level quantum system
was carried out by Autler and Townes {[}\ref{Autler}{], who making use of
Floquet's theorem derived a general solution in terms of infinite continued
fractions. These authors} applied their formulation to investigate the
effect of an rf field on the $J=2\rightarrow 1\;l$-type doublet microwave
absorption lines of molecules of gaseous OCS, obtaining good agreement with
the experimental results. The practical usefulness of their analytical
treatment decreases, however, as the oscillating field becomes stronger.

In another significant paper Shirley [\ref{Shir1}] proposed the Floquet
theory as a convenient formalism for treating periodically driven quantum
systems in the semiclassical approximation for the driving field. He then
made use of this formalism (which replaces the time-dependent Hamiltonian
with a time-independent Hamiltonian represented by an infinite matrix) to
obtain, in particular, closed expressions for time-average resonance
transition probabilities of a strongly-driven two-level system. Other works
in the same spirit are those by Ritus [\ref{Ritus1}] and Zeldovich [\ref
{Zeldo1}]. The Floquet formalism has been further elaborated in many
subsequent papers [\ref{Sambe1}--\ref{Zhao}].

A different approach, especially suited to the nonperturbative regime, was
followed by Cohen-Tannoudji and coworkers [\ref{Haro1}] to study the effects
of a nonresonant linear rf field on the Zeeman hyperfine spectra of H$^1$
and Rb$^{87}$. These authors, using the dressed atom approach [\ref{Haro2}%
] which treats the external field as a single-mode quantum field, were able
to account theoretically for the main features of the spectra, obtaining, in
particular, the collapse of the spectral lines at the zeros of the proper
zeroth-order Bessel function{. }

More recently, the driven two-level system has been used to study other
interesting phenomena. For instance, this model can properly describe the
effect of a driving laser field on the tunneling dynamics of low-lying
electrons in a double quantum well [\ref{Holt1}]. In this regard, it has
been shown [\ref{Jgom1},\ref{Gros2}] that an analytical solution which is
zeroth order in the small parameter $\Delta _0/\omega _{{\rm L}}$ accounts
correctly, over a wide parameter range, for the coherent suppression of
tunneling that occurs in driven symmetric double-well potentials {[}\ref
{Grif1}--\ref{Met2}{]}. On the other hand, because of the fact that strongly
driven two-level systems already display the main features of the high-order
harmonic generation observed experimentally in the emission spectrum of
atoms in very intense laser fields {[}\ref{Sund1}], this simple model has
also been extensively used recently {[}\ref{Yu1}--\ref{Garra1}] to
understand the basic mechanism underlying such a phenomenon [\ref{nota}]. In
particular, first-order perturbative solutions have been derived for the
equation of motion governing the dynamical evolution of the induced dipole
moment {[}\ref{Yu1}--\ref{Dak1}{]}. A different approach has been followed
in Ref. {[}\ref{VD1}] to obtain a general first-order perturbative
expression for the system time-dependent density operator, which is
applicable regardless of the coupling strength value.

In this work we are interested in a complete analytical description in the
energy domain of two-level quantum systems driven by far-off-resonance ($%
\Delta _0/\omega _{{\rm {L}}}\ll 1$) and/or strong laser fields ($\Omega
_0/\omega _{{\rm {L}}}\gg 1$). As already mentioned, this regime, which
corresponds to a parameter region where the semiclassical Jaynes-Cummings
model is not applicable, has proved to be relevant in situations such as
localization in a quantum double well or high-order harmonic generation in
atomic and molecular species. None of the analytical treatments carried out
so far in two-level systems{\large {\bf \ }}driven by nonresonant and/or
strong laser fields have arrived at such complete description, which not
only requires the knowledge of the quasienergy-level diagram but also of the 
{\em relevant} Floquet states. In this regime the usual perturbative methods
are no longer useful and one has to resort to a nonperturbative approach.
Here we will develop a nonperturbative approach which is in the spirit of
that by Cohen-Tannoudji and co-workers [\ref{Haro1},\ref{Haro2}], but is 
based on the more manageable Floquet theory and, unlike previous 
approaches, goes up to first order in a proper perturbative expansion 
(associated with a transformed Hamiltonian). This order of approximation 
turns out to be the one required to obtain a complete account of the  
spectroscopic properties of the system. Our approach starts with a 
convenient unitary transformation and leads to nonperturbative closed 
analytical expressions for the Floquet states, which contain {\em all} the 
relevant physical information on this kind of systems. From these expressions we can obtain 
not only the selection rules (which in fact only depend on the system 
symmetry properties) but also analytical expressions for the intensities 
of the different spectral lines. 
Our formulation completes, in a sense, the analytical treatment of driven 
two-level systems by extending its range of applicability to include 
particular situations which correspond to a parameter region where the 
semiclassical Jaynes-Cummings model is not applicable.

In Sec. II we obtain the quasienergies and semiclassical dressed states
(Floquet states) of a two-level quantum system driven by a nonresonant
and/or strong laser field in a coherent state. Then, in Sec. III the system
spectroscopic properties are analyzed in terms of its semiclassical dressed
states and corresponding energy-level diagrams. In particular, the
modifications induced in the system spectrum by the coupling with the laser
field can be easily understood within this formalism. Finally, the main
conclusions are summarized in Sec. IV.

\vspace{1.4 cm} 
%\newpage

\begin{center}
{\large {\bf {II. SEMICLASSICAL DRESSED STATES\vspace{.1 cm}\\}}}
\end{center}

Consider a two-level system driven by a laser field in a coherent state.
Under these circumstances the radiation field, which is assumed to be
linearly polarized along the same direction as the system dipole operator,
can be modeled by a classical electric field of frequency $\omega _{{\rm {L}}%
}$ and amplitude ${\bf E}_0$. The corresponding dimensionless Hamiltonian
(in units of $\hbar \omega _{{\rm {L}}}$) is thus given, in the dipole
approximation, by

\begin{equation}
\label{ec2.1}H=\frac{\Delta _0}{2\omega _{{\rm {L}}}} \left(
\sigma_{22}-\sigma _{11}\right) -\frac{\Omega _0} {\omega _{{\rm {L}}}}\cos
\left(\tau \right) \left( \sigma _{12}+\sigma _{21}\right) , 
\end{equation}
where $\Delta _0$ denotes the transition frequency between the excited state 
$|2\rangle $ and the ground state $|1\rangle $; $\sigma _{ij}\equiv
|i\rangle \langle j|$ is the system transition operator; $\Omega _0\equiv
E_0\mu /\hbar $ is the Rabi frequency, where $\mu $ (which is assumed to be
real) denotes the dipole matrix element between $|1\rangle $ and $|2\rangle$%
; and $\tau =\omega _{{\rm {L}}}t$ is a dimensionless time variable.

This system is periodic in $\tau $ with the period $T=2\pi $ of the laser
field. According to the Floquet theorem, the invariance of the Hamiltonian
under the discrete time translation $\tau \rightarrow \tau +T$ guarantees
that the general solution of the corresponding time-dependent Schr\"odinger
equation can be expressed as a linear superposition of solutions of the type
[\ref{Shir1}--\ref{Fain1}]

\begin{equation}
\label{ec2.2}|\Psi _j(\tau )\rangle =e^{-i\varepsilon _j\tau }
|\varphi_j(\tau )\rangle ,\;\;\;\;\;\;(j=1,2), 
\end{equation}
where $\varepsilon _j$ are the (dimensionless) quasienergies and $%
|\varphi_j(\tau )\rangle =|\varphi _j(\tau +T)\rangle $ are the periodic
Floquet states, which satisfy the eigenvalue equation

\begin{equation}
\label{ec2.3}\left( H-i\frac \partial {\partial \tau }\right)
|\varphi_j(\tau )\rangle =\varepsilon _j|\varphi _j(\tau )\rangle . 
\end{equation}
As can be easily seen, if $|\varphi _j(\tau )\rangle $ is a solution of 
Eq. (\ref{ec2.3}) with quasienergy $\varepsilon _j$ then

\begin{equation}
\label{ec2.4}|\varphi _j(\tau ),n\rangle \equiv e^{in\tau }|\varphi _j(\tau
)\rangle , 
\end{equation}
with $n$ being an arbitrary integer, is also a solution with quasienergy $%
\varepsilon _j+n$. The quasienergies are thus only defined mod $1$ (in units
of $\hbar \omega _{{\rm {L}}}$).

Floquet states and quasienergies are the natural generalization of
stationary states and energies of conservative Hamiltonians 
[\ref{Shir1}--\ref{Fain1}]. In particular, they can be shown to be the 
semiclassical counterpart of the dressed states appearing in a fully quantum 
treatment of the system in the presence of the quantized radiation field 
[\ref{Haro2}]. This reason makes them especially good for understanding the 
properties of periodically driven quantum systems in the energy domain. 
Information on transition frequencies, selection rules, and line intensities 
can be readily obtained from the form of the Floquet states and the 
quasienergy-level diagram.

Taking into account that according to the Floquet theorem any solution of
the time-dependent Schr\"odinger equation can be written as

\begin{equation}
\label{ec2.10}|\Psi (\tau )\rangle =\sum_je^{-i\varepsilon _j\tau }\langle
\varphi _j(0)|\Psi (0)\rangle \,|\varphi _j(\tau )\rangle , 
\end{equation}
and using the fact that $|\Psi (\tau )\rangle $ is given in terms of the
evolution operator $U(\tau ,0)$ by $|\Psi (\tau )\rangle =U(\tau ,0)|\Psi
(0)\rangle $, one immediately finds the following expression for the
spectral decomposition of the one-period propagator 
[\ref{Shir1}--\ref{Fain1}],

\begin{equation}
\label{ec2.12}U(T,0)=\sum_je^{-i\varepsilon _jT}|\varphi _j(T)\rangle
\langle \varphi _j(T)|, 
\end{equation}
where we have used the periodicity of $|\varphi _j(\tau )\rangle $ to
identify $|\varphi _j(0)\rangle $ with $|\varphi _j(T)\rangle $. As Eq. (\ref
{ec2.12}) shows, the diagonalization of $U(T,0)$ provides the quasienergies
and corresponding semiclassical dressed states at integer multiples of the
laser period $T$. On the other hand, from the periodicity of the Floquet
states and the relationship

\begin{equation}
\label{ec2.13}|\varphi _j(\tau )\rangle =e^{i\varepsilon _j\tau }
U(\tau,0)|\varphi _j(0)\rangle , 
\end{equation}
which is a direct consequence of Eq. (\ref{ec2.2}), it follows that in order
to gain a complete knowledge of the Floquet states at any time it suffices
to determine the evolution propagator in the time interval $0<\tau \leq T$.

As already mentioned, in this paper we are mainly interested in the analysis 
of the spectroscopic properties of a two-level system driven by a 
far-off-resonance ($\Delta _0/\omega _{{\rm {L}}}\ll 1$) and/or strong laser 
field ($\Omega _0/\omega _{{\rm {L}}}\gg 1$) in terms of its semiclassical
dressed states and quasienergy-level diagram. 
To this end we start by performing the following unitary transformation

\begin{equation}
\label{ec2.14}U_0(\tau )=e^{-i\phi (\tau )\left( \sigma _{12}+
\sigma_{21}\right) }, 
\end{equation}

\begin{equation}
\label{ec2.15}\phi (\tau )=\frac{\Omega _0}{\omega _{{\rm {L}}}} \sin
\left(\tau \right) , 
\end{equation}
which enables us to reformulate the problem in terms of a convenient small
Hamiltonian, proportional to $\Delta _0/\omega _{{\rm {L}}}$. Indeed, using
that

\begin{equation}
\label{ec2.16}U_0(\tau )|1\rangle =\cos \phi (\tau )|1\rangle -i\sin
\phi(\tau )|2\rangle , 
\end{equation}

\begin{equation}
\label{ec2.17}U_0(\tau )|2\rangle =-i\sin \phi (\tau )|1\rangle + \cos
\phi(\tau )|2\rangle , 
\end{equation}
one finds that the transformed Hamiltonian $H^{\prime }=U_0HU_0^{+}-
iU_0\dot U_0^{+}$ takes the form

\begin{equation}
\label{ec2.18}H^{\prime }=-\frac{\Delta _0}{2\omega _{{\rm {L}}}}
\left\{\left[ J_0\left( \zeta \right) +\alpha (\tau )\right] \left(
\sigma_{11}-\sigma _{22}\right) +\beta (\tau )\left( \sigma _{12}-
\sigma_{21}\right) \right\} , 
\end{equation}
where

\begin{equation}
\label{ec2.19}\alpha (\tau )\equiv \cos \left[ 2\phi (\tau )\right]-
J_0\left( \zeta \right) =2\sum_{n=1}^{+\infty }J_{2n}\left( \zeta \right)
\cos \left( 2n\tau \right) , 
\end{equation}

\begin{equation}
\label{ec2.20}\beta (\tau )\equiv i\sin \left[ 2\phi (\tau )\right]
=2i\sum_{n=0}^{+\infty }J_{2n+1}\left( \zeta \right) \sin \left[ (2n+1)
\tau\right] , 
\end{equation}

\begin{equation}
\label{ec2.21}\zeta \equiv \frac{2\Omega _0}{\omega _{{\rm {L}}}}, 
\end{equation}
The Bessel functions $J_n\left( \zeta \right) $ entering the right hand side
of Eqs. (\ref{ec2.19}) and (\ref{ec2.20}) come from the Fourier series
expansion of the periodic coefficients $\alpha (\tau )$ and $\beta (\tau )$.

In the far-off-resonance regime the oscillating part of the above
Hamiltonian can be considered as a small perturbation. In order to obtain a
first-order expression for the system evolution operator it is most
convenient to work in the interaction representation with respect to the
zeroth-order Hamiltonian

\begin{equation}
\label{ec2.22}H_0^{\prime }=-\frac{\Delta _0}{2\omega _{{\rm {L}}}}J_0
\left(\zeta \right) \left( \sigma _{11}-\sigma _{22}\right) . 
\end{equation}
In this representation the evolution operator $U_{{\rm I}}^{\prime }
(\tau,0)=e^{iH_0^{\prime }\tau }U^{\prime }(\tau ,0)$ satisfies the equation
of motion

\begin{equation}
\label{ec2.23}i\frac d{d\tau }U_{{\rm I}}^{\prime }(\tau ,0)=\Delta H_{{\rm I%
}}^{\prime }\,U_{{\rm I}}^{\prime }(\tau ,0), 
\end{equation}
where the perturbation $\Delta H_{{\rm I}}^{\prime }$ now reads

\begin{equation}
\label{ec2.24}\Delta H_{{\rm I}}^{\prime }\equiv -\frac{\Delta _0} {2\omega
_{{\rm {L}}}}\left[ \alpha (\tau )\left( \sigma _{11}- \sigma
_{22}\right)+\beta (\tau )\left( e^{-i\frac{\Delta _0} {\omega _{{\rm {L}}%
}}J_0\left(\zeta \right) \tau }\sigma _{12}- e^{i\frac{\Delta _0}{\omega _{%
{\rm {L}}}}J_0\left( \zeta \right) \tau }\sigma _{21}\right) \right] . 
\end{equation}
The solution of Eq. (\ref{ec2.23}) can be formally written as the following
infinite series

\begin{equation}
\label{ec2.25}U_{{\rm I}}^{\prime }(\tau ,0)={\bf 1}-i\int_0^\tau d
\tau_1\Delta H_{{\rm I}}^{\prime }(\tau _1)+(-i)^2\int_0^\tau d
\tau_1\int_0^{\tau _1}d\tau _2\Delta H_{{\rm I}}^{\prime }(\tau _1) \Delta
H_{{\rm I}}^{\prime }(\tau _2)+\ldots 
\end{equation}
In principle, this is a perturbative expansion in the small parameter $%
\Delta _0/\omega _{{\rm {L}}}$. However, from the asymptotic behavior of the
Bessel functions in the strong-field regime ($\zeta \equiv 2\Omega
_0/\omega_{{\rm {L}}}\gg 1$), i.e. [\ref{Abram}],

\begin{equation}
\label{ec2.26}J_n\left( \zeta \right) \sim \left\{ 
\begin{array}{ll}
\sqrt{2/\pi \zeta }\cos \left( \zeta -n\pi /2-\pi /4\right) & 
\mbox{    if   $n < \zeta$} \\ \left( e\zeta /2n\right)^n / \sqrt{2\pi n} & 
\mbox{    if   $n > \zeta$} 
\end{array}
\right. 
\end{equation}
it is not hard to see that, in this regime, the perturbation $\Delta H_{{\rm %
I}}^{\prime }(\tau )$ becomes of the order of $\Delta _0/ \sqrt{\omega _{%
{\rm {L}}}\Omega _0}$. Consequently, our perturbative analysis turns out
to be valid not only in the far-off-resonance regime ($\Delta _0/\omega _{%
{\rm {L}}}\ll 1$), but also in the regime where

\begin{equation}
\label{ec2.27}\Delta _0/\omega _{{\rm {L}}}\ll \sqrt{\Omega _0/\omega _{{\rm 
{L}}}}\gg 1.
\end{equation}
This means that in the case of intense driving fields, the present
formalism, which is generally applicable under nonresonant conditions, can
also describe the evolution of the quantum system under resonant or
quasi-resonant conditions. Put another way, our perturbative analysis
permits one to study analytically the behaviour of the driven system in a 
region of the parameter space which is beyond the range of applicability of 
the semiclassical Jaynes-Cummings model and where interesting phenomena such 
as coherent suppression of tunneling and high-order harmonic generation take
place.

The evolution operator $U(\tau ,0)$ associated with the Hamiltonian (\ref
{ec2.1}) can be written in terms of $U_{{\rm I}}^{\prime } (\tau ,0)$ as

\begin{equation}
\label{ec2.28}U(\tau ,0)=U_0^{+}(\tau )U^{\prime }(\tau,0)
U_0(0)=U_0^{+}(\tau )e^{-iH_0^{\prime }\tau }U_{{\rm I}}^{\prime } (\tau,0). 
\end{equation}
Substituting Eq. (\ref{ec2.24}) into Eq. (\ref{ec2.25}) and the latter into
Eq. (\ref{ec2.28}), one obtains, after performing the corresponding
integration, the following first-order perturbative expression for the
evolution operator

\begin{equation}
\label{ec2.29}U(\tau ,0)=U_0^{+}(\tau )e^{-iH_0^{\prime }\tau } \left\{ {\bf %
1}+i\frac{\Delta _0}{\omega _{{\rm {L}}}}\left[ \xi _{{\rm s}}\left(
\tau\right) \left( \sigma _{11}-\sigma _{22}\right) + \eta (\tau
)\sigma_{12}+\eta ^{*}(\tau )\sigma _{21}\right] + O\left( \epsilon
^2\right) \right\} , 
\end{equation}
where

\begin{equation}
\label{ec2.30a}\xi _{{\rm s}}\left( \tau \right) \equiv
\sum_{n=1}^{+\infty}J_{2n}\left( \zeta \right) \frac{\sin \left( 2n\tau
\right) }{2n}, 
\end{equation}

\begin{equation}
\label{ec2.30b}\xi _{{\rm a}}\left( \tau \right) \equiv
\sum_{n=0}^{+\infty}J_{2n+1}\left( \zeta \right) \frac{\cos \left[ \left(
2n+1\right) \tau \right] }{2n+1}, 
\end{equation}

\begin{equation}
\label{ec2.31}\eta (\tau )\equiv i\left[ \xi _{{\rm a}}\left( 0\right) -
e^{-i\frac{\Delta _0}{\omega _{{\rm {L}}}}J_0\left( \zeta \right) \tau } \xi
_{{\rm a}}\left( \tau \right) \right] . 
\end{equation}
The s and a subscripts in the above formulae refer to the symmetrical or
antisymmetrical behaviour of the corresponding factor under the time
translation $\tau \rightarrow \tau +\pi $, and $\epsilon $ denotes the
appropriate expansion parameter.

A similar expression has been derived by Frasca [\ref{Fras1}] using a
different approach, valid in the strong-field limit, based on a series
expansion dual to the Dyson series. The divergent secular terms appearing in
such an approach were resummed by using renormalization group methods [\ref
{Fras2}]. The approach followed in this paper, which is free of
secular terms, demonstrates that the first-order evolution operator (\ref
{ec2.29}) is also generally applicable under nonresonant conditions. It is
important to note, however, that our formula differs from Frasca's by the
factor $\exp \{-i[\Delta _0J_0(\zeta )/\omega _{{\rm {L}}}]\tau \}$ entering
Eq. (\ref{ec2.31}), which is absent in Ref. [\ref{Fras1}] and, as we shall
see, proves to be essential in the subsequent derivation of the Floquet
states.

In order to facilitate later manipulations it is convenient to rewrite $%
U(\tau ,0)$ as the following unitary operator

\begin{equation}
\label{ec2.32}U(\tau ,0)=U_0^{+}(\tau )e^{-iH_0^{\prime }\tau }\exp \left\{ i%
\frac{\Delta _0}{\omega _{{\rm {L}}}}\left[ \xi _{{\rm s}} \left(
\tau\right) \left( \sigma _{11}-\sigma _{22}\right) + \eta (\tau
)\sigma_{12}+\eta ^{*}(\tau )\sigma _{21}\right] \right\} \left[ 1+O\left(
\epsilon^2\right) \right] , 
\end{equation}
where the equality only holds up to first order in $\epsilon $. The
one-period propagator thus takes the form

\begin{equation}
\label{ec2.33}U(T,0)=e^{-i\frac{\Delta _0}{2\omega _{{\rm {L}}}}J_0\left(
\zeta \right) T\sigma _z}\exp \left\{ i\frac{\Delta _0} {\omega _{{\rm {L}}%
}}\left[ \eta (T)\sigma _{-}+\eta ^{*}(T)\sigma _{+} \right] \right\}
\left[1+O\left( \epsilon ^2\right) \right] , 
\end{equation}
with

\begin{equation}
\label{ec2.34}\sigma _z=\left( \sigma _{22}-\sigma _{11}\right)
,\;\;\;\;\sigma _{+}=\sigma _{21},\;\;\;\;\sigma _{-}=\sigma _{12}. 
\end{equation}
The operators $\sigma _z,\sigma _{+},\sigma _{-}$ so defined satisfy the Lie
Algebra of $SU(2)$. They are, in other words, the generators of the $SU(2)$
group. As is well known, any $SU(2)$ transformation can be expressed in
terms of such generators in the form $\exp \left[ i\left( a_z\sigma
_z+a_{+}\sigma _{+}+a_{-}\sigma _{-}\right) \right] $. In particular, it can
be proved that the one-period propagator (\ref{ec2.33}) may be written, up
to first-order in $\epsilon $, as

\begin{equation}
\label{ec2.35}U(T,0)=\exp \left\{ i\frac{\Delta _0}{\omega _{{\rm {L}}}}
\left[ -\frac{J_0\left( \zeta \right) T} {2} \sigma _z + \eta (T)\sigma _{-}
+ \eta^{*} (T) \sigma _{+} \right] \right\} \left[ 1 + O \left( \epsilon ^2
\right) \right] . 
\end{equation}
Diagonalization of this operator is now straightforward and leads to the
following Floquet states at $\tau =T$

\begin{equation}
\label{ec2.36}|\varphi _1(T)\rangle = |1\rangle - \frac{\Delta _0}{\omega _{%
{\rm {L}}}}\sum_{n=0}^{+\infty } \frac{J_{2n+1}\left( \zeta \right) }{2n+1}%
|2\rangle + O \left( \epsilon ^2 \right) , 
\end{equation}

\begin{equation}
\label{ec2.37}|\varphi _2(T)\rangle = \frac{\Delta _0}{\omega _{{\rm {L}}}}
\sum_{n=0}^{+\infty } \frac{J_{2n+1} \left( \zeta \right) }{2n+1} | 1
\rangle + | 2 \rangle + O \left( \epsilon ^2 \right) , 
\end{equation}
with corresponding quasienergies given by

\begin{equation}
\label{ec2.38}\varepsilon _1 = -\frac{\Delta _0}{2\omega _{{\rm {L}}}} J_0
\left( \zeta \right) \left[ 1 + O \left( \epsilon ^2 \right) \right] 
\end{equation}

\begin{equation}
\label{ec2.39}\varepsilon _2 = +\frac{\Delta _0}{2\omega _{{\rm {L}}}} J_0
\left( \zeta \right) \left[ 1 + O \left( \epsilon ^2 \right) \right] 
\end{equation}

It should be noticed that had we neglected the factor $\exp \{ -i [ \Delta_0
J_0 (\zeta ) / \omega _{{\rm {L}}}] \tau \}$ entering Eq. (\ref{ec2.31}),
the corresponding one-period propagator would not differ from the
zeroth-order expression $\exp \left( -iH_0^{\prime }T\right) $ and,
consequently, after diagonalization we would obtain $|1\rangle $ and $%
|2\rangle $ as the Floquet states at $\tau =T$. Retaining such a factor is,
therefore, essential in order to obtain a correct result.

The final expression for the semiclassical dressed states at any time $\tau$
can be derived from Eq. (\ref{ec2.13}). Using the evolution operator given
in Eq. (\ref{ec2.29}) as well as the initial form of the Floquet states
given in Eqs. (\ref{ec2.36}) and (\ref{ec2.37}), and retaining only terms up
to first order in the expansion parameter one obtains, after some algebra,

%%%%%%%%%%%%%% Fig. 1 %%%%%%%%%%%%%%%

\begin{equation}
\label{ec2.40}
\begin{array}{c}
|\varphi _1(\tau )\rangle =\left\{ \cos \phi (\tau )+i 
\frac{\Delta _0}{\omega _{{\rm {L}}}} \left[ \xi _{{\rm s}}\left(
\tau\right) \cos \phi (\tau ) - \xi _{{\rm a}}\left( \tau \right) \sin \phi
(\tau) \right] \right\} | 1 \rangle + \\ i \left\{ \sin \phi (\tau ) + i 
\frac{\Delta _0}{\omega _{{\rm {L}}}} \left[ \xi _{{\rm s}}\left( \tau
\right) \sin \phi (\tau ) + \xi _{{\rm a}} \left( \tau \right) \cos \phi
(\tau ) \right] \right\}| 2 \rangle + O \left( \epsilon ^2 \right) , 
\end{array}
\end{equation}

\begin{equation}
\label{ec2.41}
\begin{array}{c}
|\varphi _2(\tau )\rangle = i \left\{ \sin \phi (\tau )-i 
\frac{\Delta _0}{\omega _{{\rm {L}}}} \left[ \xi _{{\rm s}} \left( \tau
\right) \sin \phi (\tau ) + \xi _{{\rm a}} \left( \tau \right) \cos \phi
(\tau) \right] \right\} |1\rangle + \\ \left\{ \cos \phi (\tau ) - i \frac{%
\Delta _0}{\omega _{{\rm {L}}}} \left[ \xi _{{\rm s}} \left( \tau \right)
\cos \phi (\tau) - \xi _{{\rm a}} \left( \tau \right) \sin \phi (\tau )
\right] \right\} | 2 \rangle + O \left( \epsilon ^2 \right) . 
\end{array}
\end{equation}
These states are the first-order periodic Floquet states satisfying Eq. (\ref
{ec2.3}) with corresponding quasienergies given by Eqs. (\ref{ec2.38}) and (%
\ref{ec2.39}). This is the central result of this paper. All the
relevant spectroscopic and dynamical properties that driven two-level
quantum systems exhibit in the parameter range considered in this work can
be inferred from the above analytical expressions. In other words, the
Floquet states at first order already contain all the relevant physical
information on the system properties in the region of parameter space where
conditions $\Delta _0 / \omega _{{\rm {L}}} \ll 1$ or $\Delta _0 / \omega _{%
{\rm {L}}} \ll \sqrt{\Omega _0 / \omega _{{\rm {L}}}} \gg 1$ hold.

As the above expressions reflect, both $|\varphi _1(\tau )\rangle $ and $%
|\varphi _2 (\tau) \rangle $ are linear superpositions of the undriven
atomic states $|1\rangle $ and $|2\rangle $ with oscillating weights whose
mean values are basically determined by the dimensionless coupling parameter 
$\zeta$. For small $\zeta$ each Floquet state is dominated by a single
atomic state (see Fig. 1a). The contribution of the other atomic state
increases with $\zeta$ (Fig. 1b), reaching its maximum value at $\zeta = \pi$
(Fig. 1c). Beyond this point the weights of the two atomic states keep
oscillating between the extreme values $0$ and $1$.

%%%%%%%%%%%%%% Fig. 1 %%%%%%%%%%%%%%%

\begin{figure}{\par\centering \resizebox{8.2cm}{8.cm}{\rotatebox{0}
{\includegraphics{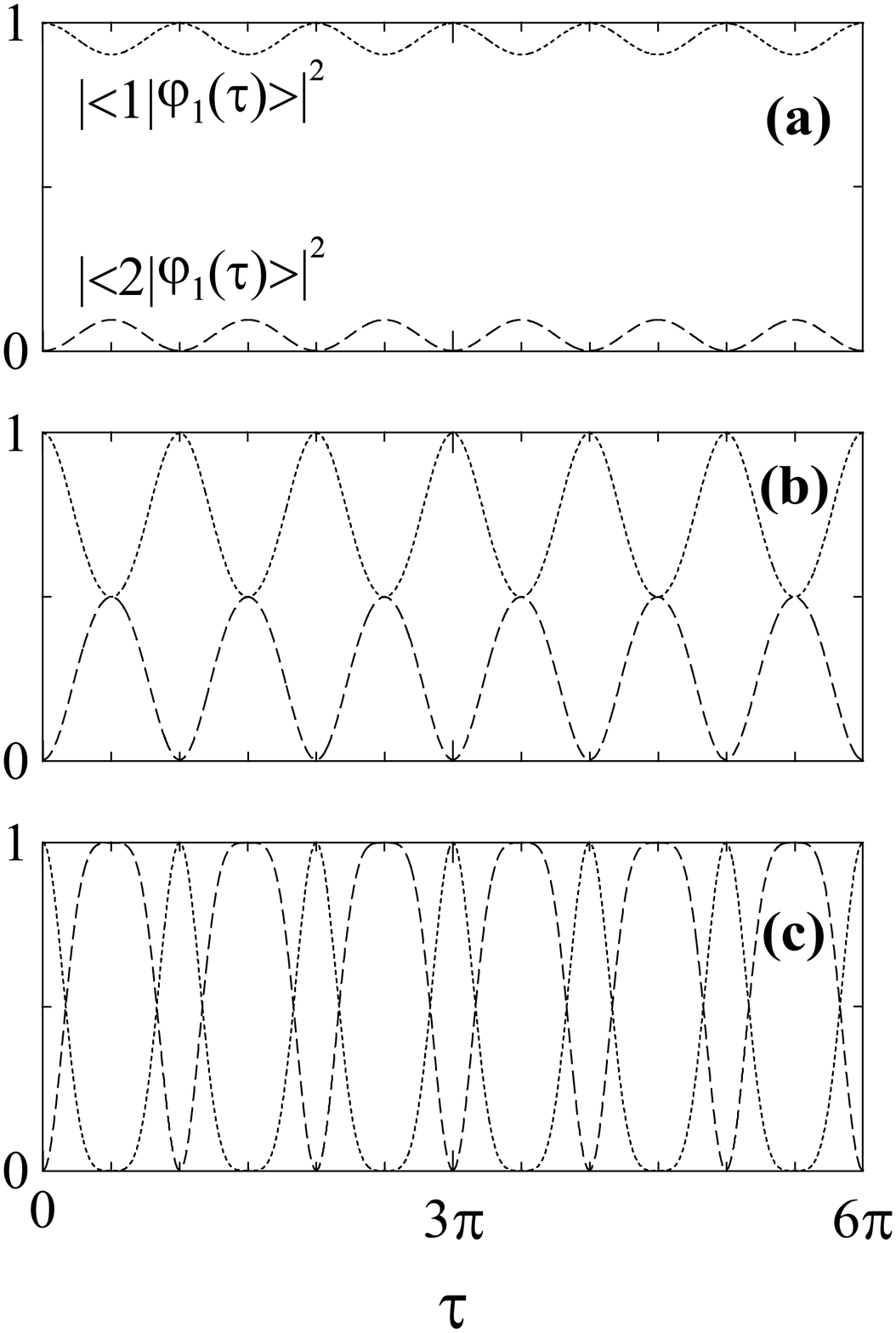}}} \par}\caption{\small
Weights of the undriven states $| 1 \rangle$ (dotted curves) and 
$| 2\rangle$ (broken curves) in the Floquet state 
$| \varphi _1 ( \tau ) \rangle$ as a function of $\tau$, for 
${\Delta _0}/{\omega _{{\rm {L}}}} = 0.1$; 
(a) $\zeta = \pi /5$; (b) $\zeta = \pi /2$; (c) $\zeta = \pi$. The 
corresponding weights in the Floquet state $| \varphi _2 (\tau ) \rangle$ 
are obtained from the above ones by interchanging the role of $|1\rangle$ 
and $|2\rangle$.
}\end{figure}

\vspace{1.4 cm}
\newpage

\begin{center}
{\large {\bf {III. FLOQUET STATE SPECTROSCOPY\vspace{.1 cm}\\}}}
\end{center}

As already mentioned, Floquet states and quasienergies are
the natural generalization of stationary states and energies of conservative
Hamiltonians. In particular, they can be shown to correspond to the dressed
states appearing in a fully quantum treatment of the system in the presence
of the radiation field. This makes them particularly convenient to
study the spectroscopic properties of a system driven by a classical
external field.

The physical interpretation of Floquet states becomes more transparent by
introducing an extended Hilbert space [\ref{Sambe1}] consisting of all $T$%
-periodic functions $\chi _i\left( {\bf r},\tau \right) $ normalizable with
respect to the scalar product defined by

\begin{equation}
\label{ec2.5}\ll \chi _i|\chi _j\gg =\frac 1T\int_0^Td\tau \int d{\bf r\,}
\chi _i^{*}\left( {\bf r},\tau \right) \chi _j\left( {\bf r},\tau \right) . 
\end{equation}
A suitable basis in this extended Hilbert space is that composed of the
Floquet states $|j,n\rangle $ corresponding to the undriven Hamiltonian,

\begin{equation}
\label{ec2.6}| j,n \rangle \equiv e^{i n \tau } |j\rangle , 
\end{equation}
where $j=1,2$ refers to the atomic state and the Fourier index $n$ is the
semiclassical analog of the photon number appearing in a quantum treatment
of the radiation field.

One of the main advantages of the semiclassical dressed state formalism is
that the spectroscopic properties of periodically driven systems can be
inferred from matrix elements of the corresponding transition operator in a
way that closely resembles that commonly used when treating conservative
Hamiltonians. In particular, the spectral lines of the allowed dipole
transitions between different dressed states $|\varphi _j(\tau ),m\rangle $
and $|\varphi _i(\tau ),n\rangle $ have frequencies $\omega =\varepsilon
_j-\varepsilon _i+m-n$ (in units of $\omega _{{\rm {L}}}$), and the
corresponding intensities are proportional to

\begin{equation}
\label{ec2.9}I_{in,jm} \equiv \left| \ll \varphi _i (\tau ),n | \hat d |
\varphi_j (\tau ),m \gg \right| ^2 = \left| \ll \varphi _i (\tau ) | \hat d
e^{i(m-n) \tau} | \varphi _j(\tau ) \gg \right| ^2, 
\end{equation}
where $\hat d=\mu \left( \sigma _{12}+\sigma _{21}\right) $ is the dipole
operator.

%%%%%%%%%%%%%%%%% Fig. 2 %%%%%%%%%%%%%%%%%

\begin{figure}{\par\centering \resizebox*{8.2cm}{8.5cm}{\rotatebox{0}
{\includegraphics{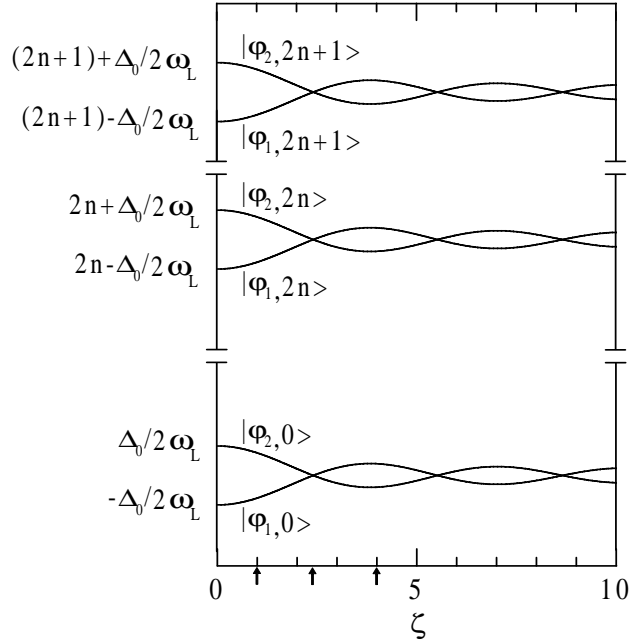}}} \par}\caption{\small 
Quasienergy-level diagram for a few n-manifolds as a function of the
dimensionless coupling parameter $\zeta$. The arrows in the $\zeta$-axis
correspond to the three different physical situations considered in Fig. 3.
}\end{figure}

In our system these transition frequencies depend on the driving field
parameters $\omega _{{\rm {L}}}$ and $\Omega _0$ entering the quasienergy
expressions (\ref{ec2.38}) and (\ref{ec2.39}). This dependence can be
observed in Fig. 2 where dressed-state levels corresponding to a few
n-manifolds are plotted against the dimensionless coupling $\zeta $. This
figure shows that level crossing occurs between the two levels in each
n-manifold whenever the driving field parameters are tuned in such a way
that the Bessel function $J_0(\zeta )$ vanishes. The existence of these
level crossings, which is possible because, as will be seen, the
corresponding states have different symmetry, leads to different
spectroscopic properties depending on whether the system is located just at
the crossing point or in either side.

The symmetry properties of the system can be used to obtain from expression (%
\ref{ec2.9}) the selection rules for the allowed transitions. In particular,
the states $|\varphi _1(\tau )\rangle $ and $|\varphi _2(\tau )\rangle $
given in Eqs. (\ref{ec2.40})--(\ref{ec2.41}) turn out to be symmetric and
antisymmetric, respectively, under the combined transformation

\begin{equation}
\label{ec2.60}|1\rangle \rightarrow |1\rangle ,\;\;\;\;\;|2\rangle
\rightarrow -|2\rangle ,\;\;\;\;\tau \rightarrow \tau +\pi , 
\end{equation}
which leaves the system Hamiltonian invariant. More generally, from this
result and the definition (\ref{ec2.4}) it follows that the Floquet states $%
|\varphi _1(\tau ),n\rangle $ and $|\varphi _2(\tau ),n+1\rangle $ are
symmetric for even $n$ and antisymmetric for odd $n$. Since the dipole
operator is antisymmetric under the above transformation, the matrix
elements (\ref{ec2.9}) between Floquet states with the same symmetry must
vanish. Therefore, the selection rules for the allowed transitions between $%
|\varphi _j(\tau ),m\rangle $ and $|\varphi _i(\tau ),n\rangle $ are

\begin{equation}
\label{ec2.61}
\begin{array}{c}
i=j,\;\;\;(m-n)\; 
{\rm odd}, \\ i\neq j,\;\;\;(m-n)\;{\rm even}. 
\end{array}
\end{equation}

Fig. 3 displays diagrams of the allowed transitions for the three different
values of the parameter $\zeta $ indicated with arrows in Fig. 2. These
values have been selected in order to explore the relevant physical
situations. In Fig. 3a the driving field does not change the level ordering
of the undriven system. As the $\zeta $ parameter increases level crossing
occurs at the first zero of the Bessel function $J_0(\zeta )$ and after
that, the original level ordering is inverted. These cases are illustrated
in Figs. 3b and 3c, respectively.

%%%%%%%%%%%%%%%% Fig. 3 %%%%%%%%%%%%%%%%%

\begin{figure}{\par\centering \resizebox*{11cm}{!}{\rotatebox{-90}
{\includegraphics{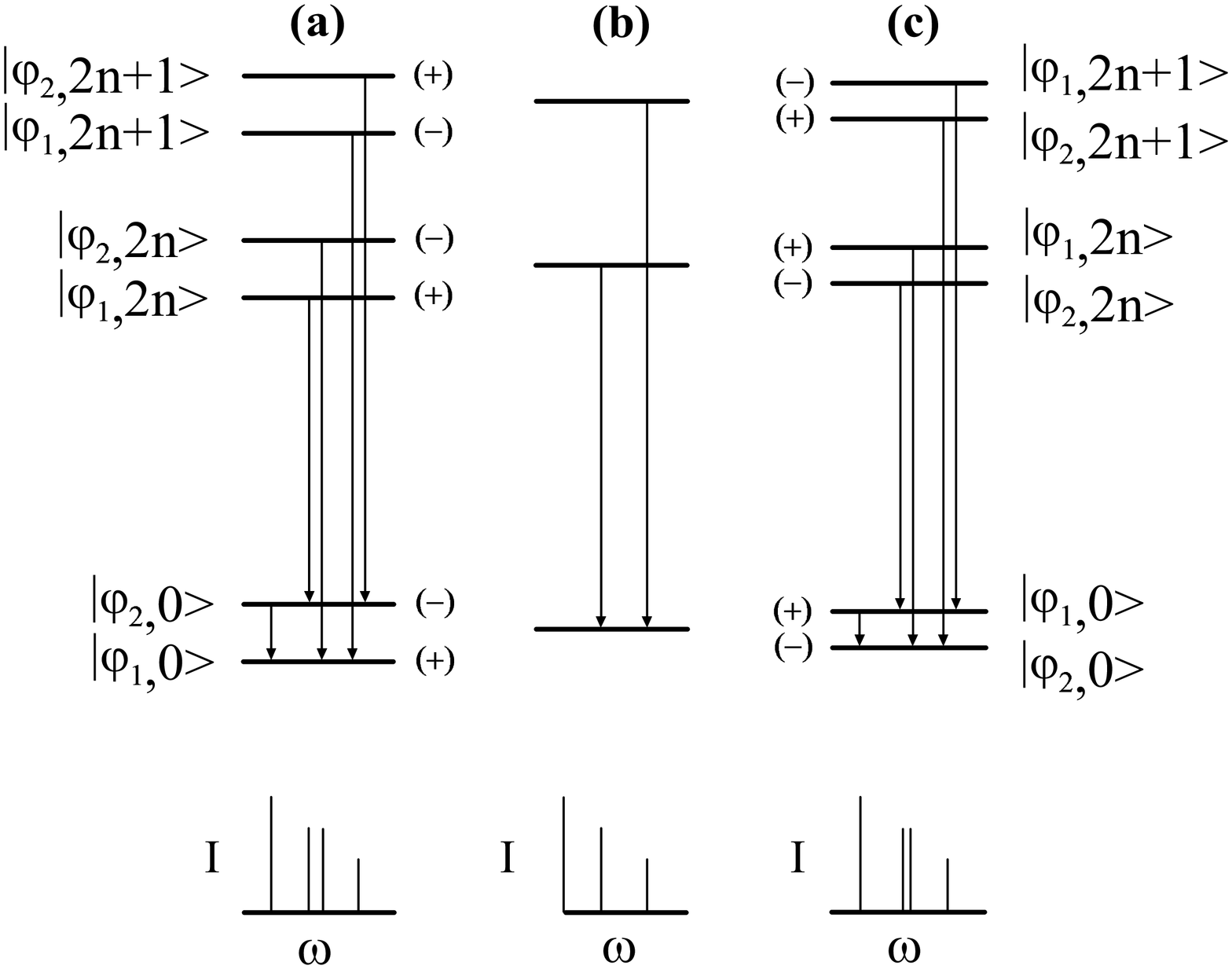}}} \par}\caption{\small 
Allowed transitions between the different semiclassical dressed states for
the three values of $\zeta$ considered in Fig. 2. The ($+$) and ($-$) symbols 
label symmetric and antisymmetric states, respectively. Only transitions
ending in the $n=0$ manifold are plotted. 
The spectra at the bottom show the qualitative intensities I of these
transitions as a function of the frequency $\omega$. Note that the most 
intense line in case (b) is a zero-frequency line corresponding to the 
transitions between the two degenerate states within the same n-manifold.
}\end{figure}

The transition between the two states within a given n-manifold ($i \neq
j,\;m=n$) gives rise to a low frequency spectral line located at $\omega =
\left| \Delta _0 J_0 (\zeta ) / \omega _{{\rm {L}}} \right| $, whose
intensity, after substitution of Eqs. (\ref{ec2.40})--(\ref{ec2.41}) into
Eq. (\ref{ec2.9}), is found to be proportional to

\begin{equation}
\label{ec2.62}I_{1n,2n}=\mu ^2 . 
\end{equation}

Note that the frequency of this transition goes to zero as the level
crossing is approached (Fig. 3b), which reflects the fact that in such a
case the dipole moment develops a constant component. On either side of 
the level crossing (Figs. 3a and 3c) the initial and final states of the
transition are interchanged. This implies, for instance, that to the right
of the first level crossing the transition initial state in the emission
spectrum would correspond to the ground state of the undriven system.

As shown in Figs. 3a and 3c, transitions between different manifolds ($m
\neq n $) with $(m-n)$ an even number, connect different atomic states ($i
\neq j$) and give rise to a series of hyper-Raman doublets with frequencies $%
\left| m-n \right| \pm \Delta _0 J_0 (\zeta ) / \omega _{{\rm {L}}}$. The
intensities of these lines, calculated as before from our Floquet states,
turn out to be proportional to

\begin{equation}
\label{ec2.63}I_{1n,2m}=\mu ^2\left( \frac{\Delta _0}{\omega _{{\rm {L}}}}%
\right) ^2\left| \frac{J_{\left| m-n\right| }(\zeta )}{m-n}\right|
^2,\;\;\;\;\;\;(m-n)\;{\rm even.} 
\end{equation}
At the zeros of $J_0(\zeta )$, that is, at the level crossings, the two
lines in each doublet collapse into a single line with twice the above
intensity (see Fig. 3b).

Finally, transitions between different manifolds with $(m-n)$ an odd number,
only take place between the same atomic states ($i=j$) and give rise to a
series of odd harmonics with frequencies $\left| m-n \right| $ and
intensities proportional to

\begin{equation}
\label{ec.64}I_{in,im}=\mu ^2\left( \frac{\Delta _0}{\omega _{{\rm {L}}}}%
\right) ^2\left| \frac{J_{\left| m-n\right| }(\zeta )}{m-n}\right|
^2,\;\;\;\;\;\;(m-n)\;{\rm odd.} 
\end{equation}

The Fourier components of the system dipole-moment expectation value are
proportional to the intensities reported here [\ref{Yu1}--\ref{Dak1}]. Our
results, based on the allowed transitions between the different
semiclassical dressed states, provide a deeper insight into the physical
origin of the spectral lines and also permit a straightforward understanding
of the driving-field influence on the spectrum features.

\vspace{1.4 cm}

\begin{center}
{\large {\bf {IV. CONCLUSION\vspace{.1 cm}\\}}}
\end{center}

The Floquet or semiclassical dressed state formalism allows periodically
driven quantum systems to be treated in a way similar to that used for
time independent systems; quasienergies and Floquet states playing a role
analogous to energies and stationary states. In particular the system
spectroscopic properties can be studied in both cases in terms of
appropriate matrix elements of the transition operator between the
corresponding different states.

In this work we have been interested in a complete analytical description in
the energy domain of driven two-level systems under nonresonant or 
strong-field conditions. These systems have been the subject of recent 
interest because they exhibit a wealth of interesting physical phenomena, 
such as coherent population trapping, localization, and high-order harmonic
generation. Under the above conditions, however, the usual perturbative
approach of the semiclassical Jaynes-Cummings model is not applicable.
Previous nonperturbative approaches do not arrive at a complete description
of the system properties in the energy domain, which requires knowledge of
the dressed states at first order in the proper small parameter.

This paper completes the analytical treatment of driven two-level
systems in the case of nonresonant and/or strong driving fields. In
particular, we have obtained analytical expressions for the system
semiclassical dressed states under these special conditions. More
specifically, by using a suitable unitary transformation we have written the
two-level Hamiltonian in a form particularly convenient for carrying out a
perturbative treatment in the small parameter $\Delta _0/\omega _{{{\rm {L}}}%
}$, with a zeroth-order Hamiltonian which includes all the secular terms. In
this way, we have obtained a first-order approximation for the evolution
operator, which turns out to be valid not only in the far-off-resonance
regime but also in the strong-field limit. Diagonalization of the one-period
evolution operator was performed with the help of $SU(2)$ algebra to obtain
the first-order Floquet states and their corresponding quasienergies.

These analytical states already account for all the relevant spectroscopic
and dynamical properties inherent to these systems in the region of
parameter space where conditions $\Delta _0/\omega _{{\rm {L}}}\ll 1$ or $%
\Delta _0/\omega _{{\rm {L}}}\ll \sqrt{\Omega _0/\omega _{{\rm {L}}}}\gg 1$
hold. In particular, these results have enabled us to understand the system
spectroscopic properties in terms of the allowed transitions between the
different semiclassical dressed states.

\vspace{1.2 cm}

\begin{center}
{\large {\bf {ACKNOWLEDGMENTS \vspace{.1 cm}\\}}}
\end{center}

This work has been supported by DGESIC (Spain) under Project No.
PB97-1479-C02-01.

%\newpage 

\vspace{1.2 cm}

\begin{center}
{\large {\bf REFERENCES \vspace{.4cm}}}
\end{center}

\begin{enumerate}
\item  \label{Jaynes} E. T. Jaynes and F. W. Cummings, Proc. IEEE {\bf 51},
89 (1963).

\item  \label{Autler} S. H. Autler and C. H. Townes, Phys. Rev. {\bf 100},
703 (1955).

\item  \label{Shir1} J. H. Shirley, Phys. Rev. {\bf 138}, B979 (1965).

\item  \label{Ritus1} V. I. Ritus, Zh. Eksp. Teor. Fiz. {\bf 51}, 1544
(1966).

\item  \label{Zeldo1} Ya B. Zeldovich, Zh. Eksp. Teor. Fiz. {\bf 51}, 1492
(1966).

\item  \label{Sambe1} H. Sambe, Phys. Rev. A {\bf 7}, 2203 (1973).

\item  \label{Fain1} A. G. Fainshtein, N. L. Manakov, and L. P. Rapoport, J.
Phys. B {\bf 11}, 2561 (1978).

\item  \label{Milfe1} S. C. Leasure, K. F. Milfeld and R. E. Wyatt, J. Chem.
Phys. {\bf 74}, 6197 (1981).

\item  \label{Milfe2} K. F. Milfeld and R. E. Wyatt, Phys. Rev. A {\bf 27},
72 (1983).

\item  \label{Zhao} X.-G. Zhao, Phys. Rev. B {\bf 49}, 16753 (1994); Phys.
Lett. A {\bf 193}, 5 (1994).

\item  \label{Haro1} S. Haroche, C. Cohen-Tannoudji, C. Audoin and J. P.
Schermann, Phys. Rev. Lett. {\bf 24}, 861 (1970).

\item  \label{Haro2} C. Cohen-Tannoudji and S. Haroche, J. Phys. (Paris) 
{\bf 30}, 153 (1969).

\item  \label{Holt1} M. Holthaus and D. Hone, Phys. Rev. B {\bf 47}, 6499
(1993).

\item  \label{Jgom1} J. M. Gomez Llorente and J. Plata, Phys. Rev. A {\bf 45}%
, R6958 (1992).

\item  \label{Gros2} F. Grossmann and P. H\"anggi, Europhys. Lett. {\bf 18},
571 (1992).

\item  \label{Grif1} For a review see, for example, M. Grifoni and P.
H\"anggi, Phys. Rep. {\bf 304}, 229 (1998).

\item  \label{Gros1} F. Grossmann, T. Dittrich, P. Jung, and P. H\"anggi,
Phys. Rev. Lett. {\bf 67}, 516 (1991).

\item  \label{Met1} R. Bavli and H. Metiu, Phys. Rev. Lett. {\bf 69}, 1986
(1992).

\item  \label{Met2} R. Bavli and H. Metiu, Phys. Rev. A {\bf 47}, 3299
(1993).

\item  \label{Sund1} B. Sundaram and P. W. Milonni, Phys. Rev. A {\bf 41},
R6571 (1990).

\item  \label{Yu1} M. Yu. Ivanov and P. B. Corkum, Phys. Rev. A {\bf 48},
580 (1993).

\item  \label{Met3} Y. Dakhnovskii and H. Metiu, Phys. Rev. A {\bf 48}, 2342
(1993).

\item  \label{Dak1}Y. Dakhnovskii and R. Bavli, Phys. Rev. B {\bf 48}, 11020
(1993).

\item  \label{Kapla1} A. E. Kaplan and P. L. Shkolnikov, Phys. Rev. A {\bf 49%
}, 1275 (1994).

\item  \label{Gaut1} F. I. Gauthey, C. H. Keitel, P. L. Knight, and A.
Maquet, Phys. Rev. A {\bf 52}, 525 (1995).

\item  \label{Pons1} M. Pons, R. Ta\"\i eb, and A. Maquet, Phys. Rev. A {\bf %
54}, 3634 (1996).

\item  \label{Gaut2} F. I. Gauthey, C. H. Keitel, P. L. Knight, and A.
Maquet, Phys. Rev. A {\bf 55}, 615 (1997).

\item  \label{Garra1} F. I. Gauthey, B. M. Garraway, and P. L. Knight, Phys.
Rev. A {\bf 56}, 3093 (1997).

\item  \label{nota} It should be noticed, however, that their physical
origins are different. Indeed, the recollision mechanism which accounts for
high harmonic generation in atoms is not possible in a two-level system, in
which ionization does not occur.

\item  \label{VD1} V. Delgado and J. M. Gomez Llorente, J. Phys. B {\bf 33},
5403 (2000).

\item  \label{Abram} M. Abramowitz and I. A. Stegun, {\it Handbook of
Mathematical Functions} (Dover, New York, 1972).

\item  \label{Fras1} M. Frasca, Phys. Rev. A {\bf 60}, 573 (1999).

\item  \label{Fras2} M. Frasca, Phys. Rev. A {\bf 56}, 1548 (1997).
\end{enumerate}

\end{document}